\documentclass[a4paper]{article}
\pdfoutput=1
 
\usepackage[margin=1.25in]{geometry}
\usepackage{verbatim}
\usepackage{amssymb,enumerate,paralist}
\usepackage{graphicx}
\usepackage{amsbsy}
\usepackage{authblk}
\usepackage{amsmath,amssymb,color}
\usepackage{graphicx}
\usepackage{amsmath,amsfonts}
\usepackage{enumerate}
\usepackage{verbatim,amssymb,amsfonts,pifont}
\usepackage{caption}

\newcommand{\D}{\mathcal{D}}

\newtheorem{theorem}{Theorem}
\newtheorem{obs}{Observation}

\newtheorem{lemma}{Lemma}
\newenvironment{proof}{\paragraph{Proof:}}{\hfill$\square$}

 \title{Exploring Increasing-Chord Paths and Trees
\thanks{Work of S. Durocher and D. Mondal is supported in part by the Natural Sciences and Engineering Research Council of Canada
 (NSERC).}
}

\author{Yeganeh Bahoo}
\author{Stephane Durocher} 
\affil[1]{Department of Computer Science, University of Manitoba, 
  Winnipeg,  Canada\\
  \texttt{\{bahoo,durocher\}@cs.umanitoba.ca}}

\author{Sahar Mehrpour}
\affil{School of Computing, University of Utah,  Utah (UT), USA\\
  \texttt{mehrpour@cs.utah.edu}}

\author{Debajyoti Mondal}
\affil{Cheriton School of Computer Science, University of Waterloo, Canada\\
  \texttt{dmondal@uwaterloo.ca}}

\begin{document}

\maketitle

\begin{abstract}
A straight-line drawing $\Gamma$ of a graph $G=(V,E)$ is a drawing of $G$ in the Euclidean plane, where every vertex in $G$ is mapped to a distinct point, and every edge in $G$ is mapped to a straight line segment between their endpoints. A path $P$ in $\Gamma$   is called  increasing-chord if for every four points (not necessarily vertices) $a,b,c,d$ on $P$  in this order, the Euclidean distance between $b,c$ is at most the Euclidean distance between $a,d$. A spanning tree $T$ rooted at some vertex $r$ in $\Gamma$ is  called increasing-chord if $T$ contains an increasing-chord path from $r$ to every vertex in $T$.   We prove that given a vertex $r$ in a straight-line drawing $\Gamma$, it is NP-complete to decide whether $\Gamma$ contains an increasing-chord spanning tree rooted at $r$, which answers a question posed by Mastakas and Symvonis~\cite{Mastakas}. We also  shed light on 
 the problem of finding an increasing-chord path between a pair of vertices in $\Gamma$, but the computational   complexity question  remains open.  
\end{abstract}

\section{Introduction}
 \label{Introduction} 
 
In 1995, Icking et al.~\cite{Klein1995} introduced the concept of a self-approaching curve. A curve is called \emph{self-approaching} if for any three points $a$, $b$ and $c$ on the curve in this order, $|bc| \leq |ac|$, where $|xy|$ denotes the Euclidean distance between $x$ and $y$. A curve is called \emph{increasing-chord} if it is self-approaching  in both directions. 
 A path $P$ in a straight-line drawing $\Gamma$ is called  increasing-chord  if for every four  points (not necessarily  vertices) $a,b,c,d$ on $P$  in this order, the inequality  $ |bc|\le |ad|$ holds. $\Gamma$ is called an \emph{increasing-chord drawing} if there exists an increasing-chord path between every pair of vertices in $\Gamma$.  

The study of increasing-chord drawings was motivated by greedy routing in geometric networks, where given two vertices $s$ and $t$, the goal is to send a message from $s$ to $t$ using some greedy strategy, i.e., at each step, the next vertex in the route is selected greedily as a function of the positions of the neighbors  of the current vertex $u$ relative to the positions of $u$, $s$, and $t$~\cite{RaoPSS03}. A polygonal path $u_1,u_2, \ldots, u_k$ is called a \emph{greedy path} if for every $i$, where $0<i<k$, the inequality $|u_iu_k| > |u_{i+1}u_k|$ holds. If a straight-line drawing is \emph{greedy}, i.e., there exists a greedy path between every pair of vertices, then it is straightforward to route the message between any pair of vertices by  following a    greedy path. For example, we  can repeatedly forward the message to some node which is closer to the destination than the current vertex. A disadvantage of a greedy drawing, however, is that the \emph{dilation}, i.e., the ratio of the graph distance to the Euclidean distance  between a pair of vertices, may  be unbounded.  Increasing-chord drawings were introduced to address this problem, where the dilation of increasing-chord drawings can be at most $2 \pi /3 \le 2.094$~\cite{Rote}. 

 Alamdari et al.~\cite{AlamdariCGLP12} examined the problem   of recognizing increasing-chord drawings, and the problem of constructing  such a drawing on a given set of points.  They showed that it is NP-hard to recognize increasing-chord drawings in $\mathbb{R}^3$, and asked whether it is also NP-hard in $\mathbb{R}^2$. They also proved that for every  set of $n$ points $P$ in $\mathbb{R}^2$, one can  construct an increasing-chord drawing $\Gamma$ with $O(n)$  vertices and edges, where $P$ is a subset of the vertices of $\Gamma$. In this case, $\Gamma$ is called a \emph{Steiner network of $P$}, and the vertices of $\Gamma$ that do not belong to $P$ are called Steiner points. Dehkordi et al.~\cite{DehkordiFG15} proved that if $P$ is a convex point set, then one can construct an increasing-chord network  with  $O(n\log n)$ edges, and without introducing any Steiner point. 
 Mastakas and Symvonis~\cite{MastakasS15} improved the $O(n\log n)$ upper bound on edges to $O(n)$ with at most one Steiner point. N{\"{o}}llenburg et al.~\cite{NollenburgPR16} examined the problem of computing increasing-chord drawings of given graphs. 
 Recently, Bonichon et al.~\cite{Bonichon2016} showed that the existence of an angle-monotone path of width $0\le \gamma<180^\circ$     between a pair of vertices (in a straight-line drawing) can be decided in polynomial time, which is very interesting since angle-monotone paths of width $\gamma\le 90^\circ$ satisfy increasing chord property.

N{\"{o}}llenburg et al.~\cite{NollenburgPR15} showed that partitioning a plane graph drawing
 into a minimum number of increasing-chord components is NP-hard, which extends a result of Tan and
Kermarrec~\cite{TanK12}. They also proved that the problem remains NP-hard for trees, and gave
 polynomial-time algorithms in some restricted settings. 
 Recently, Mastakas and Symvonis~\cite{Mastakas} showed that given a point set $S$ and a point $v\in S$,  one can compute a rooted minimum-cost spanning tree in polynomial time, where each point in $S\setminus\{v\}$ is connected to $v$ by a path that satisfies  some monotonicity property. They also proved that the existence of a monotone rooted spanning tree in a given geometric graph can be decided in polynomial time, and asked whether the decision problem remains NP-hard also for increasing-chord or self-approaching properties.

We prove that given a vertex $r$ in a straight-line drawing $\Gamma$, it is NP-complete to decide whether $\Gamma$ contains an increasing-chord spanning tree rooted at $r$, which answers the above question. We also  shed light on the problem of finding an increasing-chord path between a pair of vertices in $\Gamma$, but the computational   complexity question  remains open.  
  
\section{Technical Background}
\label{td}

\begin{figure}[pt] 
\centering
\includegraphics[width=.4\textwidth]{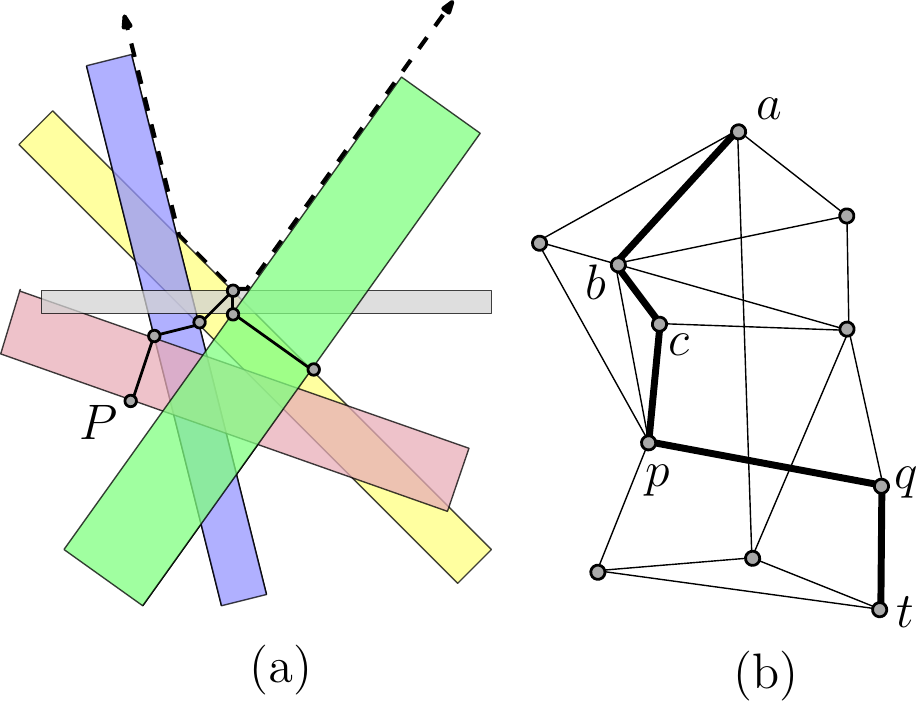} 
\caption{(a) Illustration for $\Psi(P)$, where the upper envelope is shown in dashed line. (b) An increasing-chord extension of $a,b,\ldots,p$ is shown in bold.}  
\label{technical}
\end{figure}

Given a straight line segment $l$, the \emph{slab of $l$} is an infinite region lying between a pair of parallel straight lines that are perpendicular to $l$, and pass through the endpoints of $l$.  Let $\Gamma$ be a straight-line drawing, and let $P$ be a path in $\Gamma$. Then the \emph{slabs} of $P$ are the slabs of the line segments of $P$. We denote by $\Psi(P)$ the arrangement of the slabs of $P$. Figure~\ref{technical}(a) illustrates a path $P$, where the slabs of $P$ are shown in shaded regions. Let $A$ be an arrangement of a set of straight lines such that no line in $A$ is vertical. Then the \emph{upper envelope}  of $A$ is a polygonal chain $U(A)$ such that each point of $U(A)$ belongs to some straight line of $A$, and  they are visible from the point $(0,+\infty)$. The upper envelope of a set of slabs is the upper envelope of the arrangement of lines corresponding to the slab boundaries, as shown in dashed line in Figure~\ref{technical}(a). 

Let $t$ be a vertex in $\Gamma$ and let $Q=(a,b,\ldots,p)$ be an increasing-chord path in $\Gamma$. A path $Q'=(a,b,\ldots,p,\ldots,t)$ in $\Gamma$ is called an \emph{increasing-chord extension of $Q$} if $Q'$ is also an increasing-chord path,  e.g., see Figure~\ref{technical}(b).  The following property can be derived from the definition of an increasing-chord path.


\begin{obs}[Icking et al.~\cite{icking1999self}]
\label{obs1}
A polygonal path $P$ is increasing-chord if and only if for each point $v$ on the path, the line perpendicular to $P$ at $v$ does not  properly intersect  $P$ except possibly at $v$. 
\end{obs}

A straightforward consequence of Observation~\ref{obs1} is that every polygonal chain which is both $x$- and $y$-monotone, is an increasing-chord path.  We will use Observation~\ref{obs1} throughout the paper to verify whether a path is increasing-chord.  Let $v$ be a point in $\mathbb{R}^2$. By the \emph{quadrants of $v$} we refer to the four regions determined by the vertical and horizontal lines through $v$.

\section{Increasing-Chord Rooted Spanning Trees}
\label{sec:construction}

In this section we prove the problem of computing a rooted increasing-chord
 spanning tree of a given straight-line drawing to be NP-hard. We will refer to this 
 problem as \textsc{IC-Tree}, as follows:

\bigskip
\begin{compactenum}
\item[\textbf{Problem:}] Increasing-Chord Rooted Spanning Tree \textsc{(IC-Tree)}
\item[\textbf{Instance:}] A straight-line drawing $\Gamma$ in $\mathbb{R}^2$, and a vertex $r$ in $\Gamma$.
\item[\textbf{Question:}] Determine whether $\Gamma$ contains a tree $T$ rooted at $r$ such that for each vertex $v(\not= r)$ in $\Gamma$, $T$ contains an increasing-chord path between $r$ and $v$. 
\end{compactenum}
\bigskip
\noindent
Specifically, we will prove the following theorem. 

\begin{theorem}
Given a vertex $r$ in a straight-line drawing $\Gamma$, it is NP-complete  to decide whether $\Gamma$ admits an increasing-chord spanning tree rooted at $r$. 
\end{theorem}

 We reduce the NP-complete problem \textsc{3-SAT}~\cite{Garey} to \textsc{IC-Tree}.  Let $I=(X,C)$ be an instance of \textsc{3-SAT}, where $X$ and $C$ are the set of variables and clauses. We construct a straight-line drawing $\Gamma$ and choose a vertex $r$ in $\Gamma$ such that $\Gamma$ contains an increasing-chord spanning tree rooted at $r$ if and only if $I$ admits a satisfying truth assignment.
 Here we give an outline of the hardness proof and describe the construction of $\Gamma$. 
 A detailed reduction is given in Appendix B. 

\begin{figure*}[pt] 
\centering
\includegraphics[width=.8\textwidth]{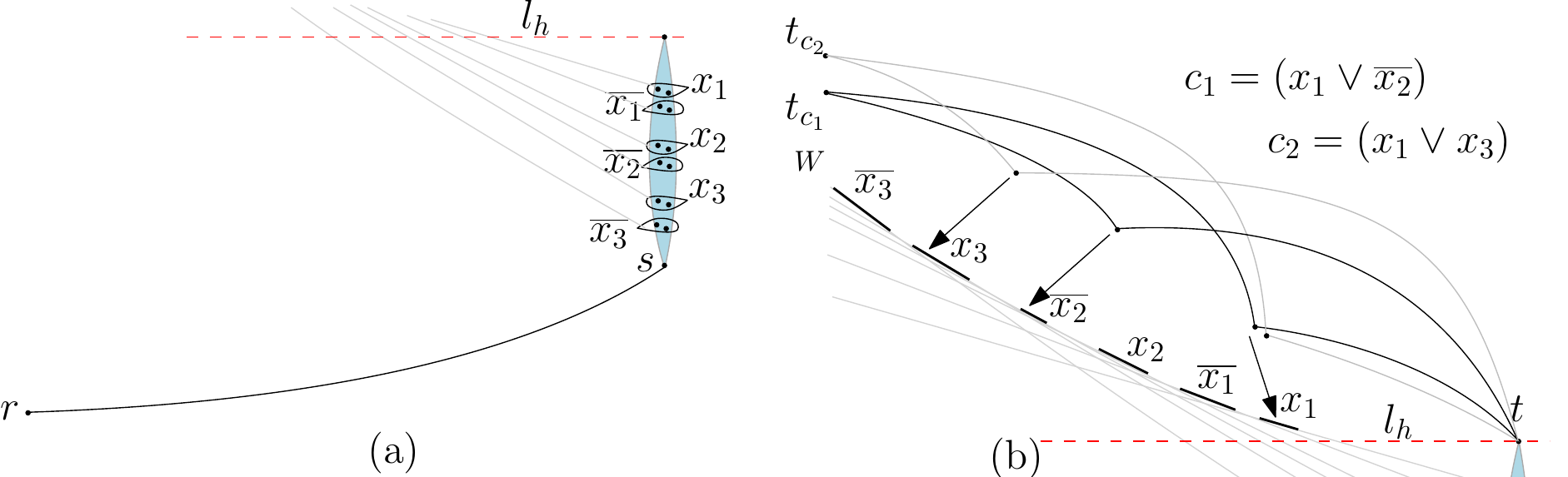} 
\caption{A schematic representation of $\Gamma$: (a) Points below $l_h$, (b) Points above $l_h$. The points that correspond to $c_1$ and $c_2$ are connected in paths of  black, and gray, respectively.  The slabs of the edges of $H$ that determine the upper envelope are shown in gray straight lines. Each variable and its negation correspond to a pair of adjacent line segments on the upper envelope of  the slabs. See Figure~\ref{idea} in Appendix A for a better illustration.
}  
\label{idea2}
\end{figure*}

Assume that $\alpha = |X|$, and $\beta = |C|$. Let $l_h$ be the line determined by the $X$-axis. $\Gamma$ will contain  $O(\beta)$ points above $l_h$, one point $t$ on $l_h$, and  $O(\alpha)$ points below $l_h$, as shown in Figures~\ref{idea2}(a)--(b).  Each clause $c\in C$ with $j$ literals, will correspond to a set of $j+1$ points above $l_h$, and we will refer to the point with the highest $y$-coordinate among these $j+1$ points as the \emph{peak} $t_c$ of $c$. Among the points below $l_h$, there are $4\alpha$ points that correspond to the variables and their negations, and  two other points, i.e., $s$  and $r$. In the reduction, the point $t$ and the points below $l_h$ altogether help to set the truth assignments of the variables.  
 
We will first create a straight-line drawing $H$ such that  every increasing-chord path between $r$ and $t_c$, where $c\in C$, passes through $s$ and $t$. Consequently, any increasing-chord tree $T$ rooted at $r$ (not necessarily spanning), which spans the points   $t_c$, must contain an increasing-chord path $P=(r,s,\ldots,t)$. We will use this path to set the truth values of the variables. 

The edges of $H$ below $l_h$ will create a set of thin slabs, and the upper envelope of these slabs will determine a convex chain $W$ above $l_h$. Each line segment on $W$ will correspond to a distinct variable, as shown in Figure~\ref{idea2}(b). 
 The points that correspond to the clauses will be positioned below these segments, and hence 
 some of these points will be `inaccessible' depending on the choice of the path $P$. 
 These literal-points will ensure that for any clause $c\in C$, there exists
 an increasing-chord extension of $P$ from $t$ to  $t_c$ if and only if $c$
 is satisfied by the truth assignment determined by $P$.

By the above discussion, $I$ admits a satisfying truth assignment if and only if there exists an increasing-chord tree $T$ in $H$ that connects the peaks to $r$. But $H$ may still contain some vertices that do not belong to this tree. Therefore, we construct the final drawing $\Gamma$ by adding some new paths to $H$, which will allow us to reach these remaining vertices  from $r$. 
 We now describe the construction in details. 

\begin{figure*}[pt] 
\centering
\includegraphics[width=.9\textwidth]{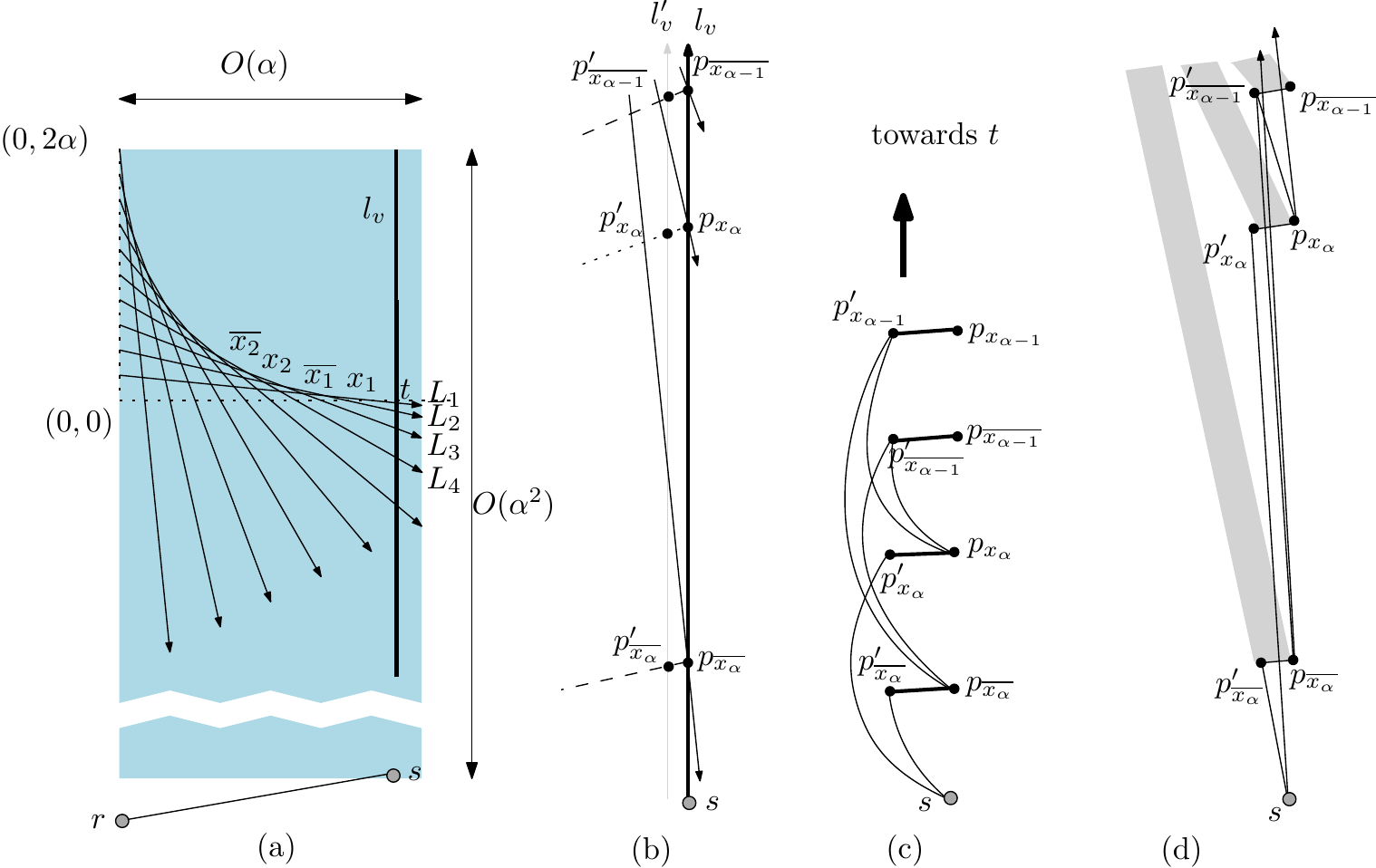} 
\caption{(a) Construction of $\mathcal{A}$. (b)--(c) Construction of the vertices and edges of $H_b$. (d)  Illustration for the  straight line segments of $H_b$, and the slabs corresponding to the needles.
}  
\label{details}
\end{figure*}

{\textbf{Construction of $H$:}}   
 We first construct an arrangement $\mathcal{A}$ of $2\alpha$ straight line segments.  The endpoints of the $i$th line segment $L_i$, where $1\le i\le 2\alpha$, are $(0,i)$ and $(2\alpha-i+1,0)$. We now extend each $L_i$ downward by scaling its length by a factor of $(2\alpha+1)$, as shown in Figure~\ref{details}(a). Later, the variable $x_j$, where $1\le j\le \alpha$, and its negation  will be represented using the lines $L_{2j-1}$ and $L_{2j}$.   
Let $l_v$ be a vertical line segment with endpoints $(2\alpha+1,2\alpha)$ and $(2\alpha+1,-5\alpha^2)$. Since the  slope of a line in $\mathcal{A}$ is in the interval $[-2\alpha, - 1/(2\alpha)]$, each $L_i$ intersects $l_v$.  Since the coordinates of the endpoints of $L_i$ and $l_v$ are of size $O(\alpha^2)$, and all the intersection points can be represented using polynomial space.
 
By construction, the line segments of $\mathcal{A}$ appear on  $U(\mathcal{A})$ in the order of the variables, i.e., the first two segments (from right) of   $U(\mathcal{A})$ correspond to $x_1$ and $\overline{x_1}$, the next two segments correspond to $x_2$ and $\overline{x_2}$, etc.

{\textbf{Variable Gadgets:}} 
 We denote the intersection point of $l_h$ and $l_v$  by $t$, and the   endpoint $(2\alpha+1, - 5\alpha^2)$ of $l_v$ by $s$. 
 We now create the points that correspond to the variables and their negations. Recall that $L_{2j-1}$ and $L_{2j}$ correspond to the variable $x_j$ and its negation $\overline{x_j}$, respectively. Denote the intersection point of $L_{2j-1}$ and $l_v$ by $p_{x_j}$, and the intersection point of $L_{2j}$ and $l_v$ by $p_{\overline{x_{j}}}$, e.g., see  Figure~\ref{details}(b). For each $p_{x_j}$
 ($p_{\overline{x_j}}$), we create a new point $p'_{x_j}$ ($p'_{\overline{x_j}}$) such that the straight line segment
 $p_{x_j}p'_{x_j}$ ($p_{\overline{x_j}}p'_{\overline{x_j}}$) is perpendicular to $L_{2j-1}$ ($L_{2j}$), as shown using the dotted (dashed) line  in Figure~\ref{details}(b). 
  We may assume that all the points $p'_{x_j}$ and $p'_{\overline{x_j}}$ lie on a vertical line $l'_v$, where $l'_v$ lies $\varepsilon$ distance away to the left of $l_v$. The value of $\varepsilon$ would be determined later. In the following we use the points $p_{x_j}$, $p_{\overline{x_j}}$, $p'_{x_j}$ and $p'_{\overline{x_j}}$ to create some polygonal paths from $s$ to $t$. 

For each $j$ from $1$ to $\alpha$,  we draw the  straight line segments $p_{{x_j}}p'_{{x_j}}$ and  $p_{\overline{x_j}}p'_{\overline{x_j}}$. Then for each $k$,  where $1< k \le \alpha $, we make   $p_{x_k}$ and $p_{\overline{x_k}}$ adjacent  to both $p'_{x_{k-1}}$ and $p'_{\overline{x_{k-1}}}$,  e.g., see Figure~\ref{details}(c). We then add the edges from $s$ to $p'_{x_\alpha}$ and $p'_{\overline{x_\alpha}}$, and finally, from $t$ to $p_{x_1}$ and $p_{\overline{x_1}}$. For each $x_j$
 ($\overline{x_j}$), we refer to the segment $p_{{x_j}}p'_{{x_j}}$ ($p_{\overline{x_j}}p'_{\overline{x_j}}$) as the \emph{needle} of $x_j$ ($\overline{x_j}$). Figure~\ref{details}(c) illustrates the needles in bold.
 Let the resulting drawing be $H_b$.

Recall that $l'_v$ is $\varepsilon$ distance away to the left of $l_v$. We choose $\varepsilon$ sufficiently small such that for each needle, its slab does not intersect any other needle in $H_b$, e.g., see Figure~\ref{details}(d).  The upper envelope of the slabs of all the straight line segments of $H_b$ coincides with $U(\mathcal{A})$. Since the distance between any pair of points that we created on $l_v$ is at least $1/\alpha$ units, it suffices to choose $\varepsilon = 1/\alpha^3$. Note that the points $p'_{x_j}$ and $p'_{\overline{x_j}}$ can be represented in polynomial space using the endpoints of $l'_v$ and the endpoints of the segments  $L_{2j-1}$ and $L_{2j}$. 
 The proof of the following lemma is included in   Appendix A.
  
\begin{lemma}
\label{lem:path}
Every increasing-chord path $P$ that starts at $s$ and ends at $t$ must pass through exactly one point among $p_{x_j}$ and $p_{\overline{x_j}}$, where $1 \le j\le \alpha$, and vice versa.
\end{lemma}

We now place a point $r$ on the $y$-axis sufficiently below $H_b$, e.g., at position $(0,-\alpha^5)$, such that the slab of the straight line segment $rs$ does not intersect $H_b$ (except at $s$), and similarly, the slabs of the line segments of $H_b$ do not intersect $rs$. 
 Furthermore, the slab of $rs$ does not intersect any segment $L_j$, and vice versa.  We then add the point $r$ and the segment $rs$ to $H_b$. Let $P$ be an increasing-chord path from $r$ to $t$. The upper envelope of $\Psi(P)$   is determined by the needles in $P$, which selects some segments from the convex chain $W$, e.g., see Figure~\ref{idea2}(b). For each $x_j$, $P$ passes through exactly one point among $p_{x_j}$ and $p_{\overline{x_j}}$. Therefore, for each variable $x_j$, either the slab of $x_j$, or the slab of $\overline{x_j}$  appears on $U(P)$.  Later, if $P$ passes through point $p_{x_j}$ ($p_{\overline{x_j}}$), then we will set $x_j$ to false (true). Since $P$ is an increasing-chord path, by Lemma~\ref{lem:path} it cannot pass through both  $p_{x_j}$ and  $p_{\overline{x_j}}$ simultaneously. Therefore, all the truth values will be set consistently.

{\textbf{Clause Gadgets:}} 
We now complete the construction of $H$ by adding clause gadgets to $H_b$. For each clause $c_i$, where $1\le i\le \beta$, we first create the peak point $t_{c_i}$ at position $(0,2\alpha+i)$. For each variable $x_j$, let $\lambda_{x_j}$  be the interval of $L_{2j-1}$ that appears  on the upper envelope of $\mathcal{A}$. Similarly, let $\lambda_{\overline{x_j}}$ be the  interval of  $L_{2j}$   on the upper envelope of $\mathcal{A}$.  For each $c_i$, we  construct a point $q_{x_j,c_i}$ ($q_{\overline{x_j},c_i}$) inside the cell of $\mathcal{A}$  immediately below   $\lambda_{x_j}$ ($\lambda_{\overline{x_j}}$).   We will refer to these points as the \emph{literal-points of   $c_i$}.  Figure~\ref{tree} in  Appendix B depicts these points in black squares. 
 We assume that for each variable, the corresponding literal-points 
 lie on the same location. One may perturb them to remove vertex overlaps. 
 For each variable $x\in c_i$, we create a path $(t, x, t_c)$.  
 In the reduction, if at least one of the literals of  $c_i$ is true, 
 then we can take the corresponding path
 to connect $t_c$ to $t$. Let the resulting drawing be $H$. 


{\textbf{Construction of $\Gamma$:}}   
 Let $q$ be a literal-point in $H$. We now add an increasing-chord path  $P' = (r,a,q)$ to $H$ in such a way that $P'$ cannot be extended to any larger increasing-chord path in $H$. We place the point $a$ at the intersection point of the horizontal line through $q$ and the  vertical line through $r$, e.g., see Figure~\ref{tree}(b) in Appendix B. We refer to the point $a$ as the \emph{anchor} of $q$.  By the construction of $H$, all the neighbors of $q$ that have a higher $y$-coordinate than $q$ lie in the top-left quadrant of $q$, as illustrated by the  dashed rectangle in Figure~\ref{tree}(b).  Let $q'$ be the first neighbor in the top-left quadrant of $q$ in counter   clockwise order. Since $\angle aqq'< 90^\circ$, $P'$ cannot be extended to any larger increasing-chord path $(r,a,q,w)$ in $H$, where the $y$-coordinate of $w$ is higher than $q$.  On the other hand, every   literal-point $w$ in $H$ with $y$-coordinate smaller than $q$ intersects the slab of $ra$. Therefore, $P'$ cannot be extended to any larger increasing-chord path. 

For every literal-point $q$ in $H$, we add such an increasing-chord path from $t$ to $q$. To avoid edge overlaps, one can  perturb the anchors such that  the new paths remain  increasing-chord and non-extensible to any larger increasing-chord paths. This completes the construction of $\Gamma$. We refer the reader to Appendix B for the  formal details of the reduction.

\begin{figure*}[pt] 
\centering
\includegraphics[width=.75\textwidth]{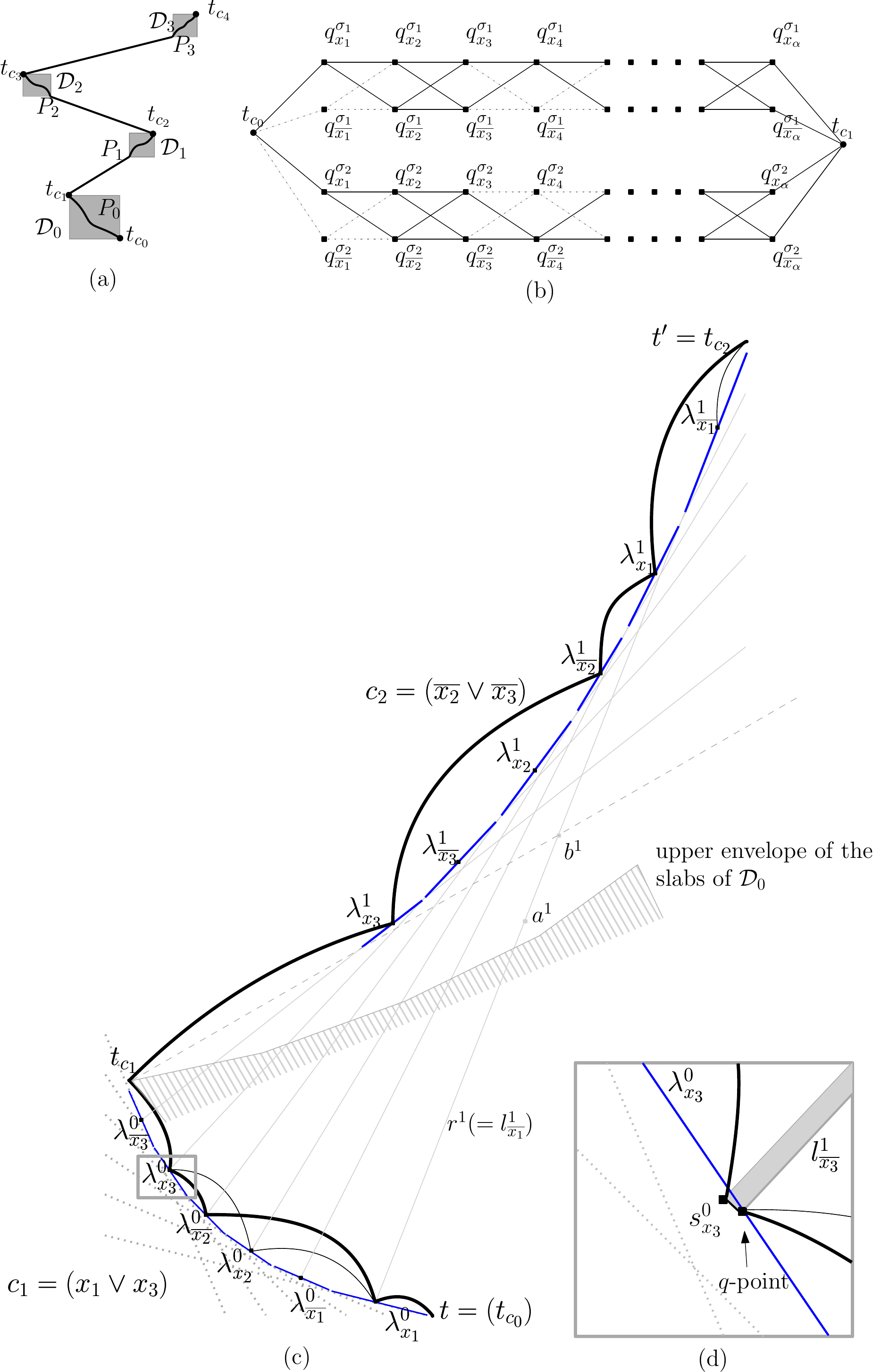} 
\caption{ (a) Idea for the reduction.  (b) The graph corresponding to the truth  assignment satisfying $c_1 = (x_1\vee x_4)$. Only the construction for the truth assignments $\sigma_1=\{x_1=true, x_4 = true\}$ and  $\sigma_2=\{x_1=true, x_4 = false\}$ are shown. 
  (c) A schematic representation for $\D$, where $I = (x_1\vee {x_3}) \wedge (\overline{x_2}\vee \overline{x_3})$. An increasing-chord path is shown in bold, and the corresponding truth value assignment is: $x_1=true$, $x_2=false$, $x_3=true$.  (d) Illustration for an $s$-segment. 
}  
\label{graph}
\end{figure*} 
 
\section{Increasing-Chord Paths}
\label{sec:path}

In this section we attempt to reduce 3-SAT to the problem of finding an increasing-chord path (\textsc{IC-Path}) between a pair of vertices in a given straight-line drawing. 
 We were unable to bound the coordinates of the drawing to a polynomial number of bits, and hence the computational complexity question of the problem remains open. We hope that the ideas we present here will be useful in future endeavors to settle the question.

  
Here we briefly describe the idea of the reduction. Given a 3-SAT instance $I=(X,C)$,  the corresponding drawing $\D$ for \textsc{IC-Path} consists of straight-line drawings $\D_{i-1}$, where $1\le i\le \beta$, e.g., see Figure~\ref{graph}(a). The drawing $\D_{i-1}$ corresponds to the each clause $c_i$. We will refer to the bottommost (topmost) point of $\D_{i-1}$ as $t_{c_{i-1}}$ ($t_{c_i}$). We will choose $t_{c_0}$ and $t_{c_\beta}$ to be the points $t$ and $t'$, respectively, and show that $I$ admits a satisfying truth assignment if and only if there exists an  increasing-chord  path $P$ from $t$ to $t'$ that passes through every  $t_{c_i}$. For every $i$, the subpath $P_{i-1}$ of $P$ between $t_{c_{i-1}}$ and  $t_{c_i}$ will correspond to a set of truth values for all the variables in $X$. The most involved part is to show that the truth values determined by $P_{i-1}$ and $P_i$ are consistent. This consistency will be ensured by the construction of $\D$, i.e.,   the increasing-chord path $P_{i-1}$ from $t_{c_{i-1}}$ to $t_{c_i}$ in $\D_{i-1}$ will determine a set of slabs, which will force a unique increasing-chord path $P_{i}$ in $\D_i$ between $t_{c_{i}}$ and $t_{c_{i+1}}$ with the same truth values as determined by $P_{i-1}$.   

{\textbf{Construction of $\D$:}} 
 The construction of $\D_{i-1}$ depends on an arrangement of lines $\mathcal{A}^{i-1}$. The construction of $\mathcal{A}^0$ is the same as the construction of arrangement $\mathcal{A}$, which we described in Section \ref{sec:construction}.  Figure~\ref{graph}(c) illustrates $\mathcal{A}^0$ in dotted lines.  For each variable $x_j$, where $1\le j \le \alpha$, there exists an interval $\lambda^0_{x_j}$ of $L_{2j-1}$ on the upper envelope of $\mathcal{A}^0$. Similarly, for each   $\overline{x_j}$,  there exists an interval $\lambda^0_{\overline{x_j}}$ of $L_{2j}$ on the upper envelope of $\mathcal{A}^0$.

 
We now describe the construction of $\D_0$. Choose $t_{c_0}$ ($t_{c_1}$) to be the bottommost (topmost)   point of $\lambda^0_{x_1}$ ($\lambda^0_{\overline{x_\alpha}}$). We then slightly shrink the intervals  
 $\lambda^0_{x_1}$ and $\lambda^0_{\overline{x_\alpha}}$ such that $t_{c_0}$ and  $t_{c_1}$ no longer belong to these segments.  Assume that $c_1$  contains $\delta$ literals, where $\delta\le 3$, and let   $\sigma_1,\ldots,\sigma_{2^\delta -1}$ be the satisfying truth assignments for $c_1$. We construct a graph $G_{c_1}$ that corresponds to these satisfying truth assignments,  e.g., see Figure~\ref{graph}(b) and Appendix C for formal details. The idea is to ensure that any path between $t_{c_0}$ and $t_{c_1}$ passes through exactly one point in $\{q^{\sigma_k}_{x_j}, q^{\sigma_k}_{\overline{x_j}}\}$, for each truth assignment $\sigma_k$, which will set the truth value of $x_j$.  In $\mathcal{D}_0$, the point  $q^{\sigma_k}_{x_j}$ ($q^{\sigma_k}_{\overline{x_j}}$) is chosen to be the midpoint of $\lambda^{i-1}_{x_j}$ ($\lambda^{i-1}_{\overline{x_j}}$). Later, we will refer to these points as \emph{$q$-points}, e.g., see Figure~\ref{graph}(c).  We may assume that for each $x_j$, the points $q^{\sigma_k}_{x_j}$ lie at the same location. At the end of the construction, one may perturb them to remove vertex overlaps.

By Observation~\ref{obs1}, any $y$-monotone path $P'$ between $t_{c_0}$ and $t_{c_1}$ must be an increasing-chord path.  
 If $P'$ passes through  $q^\sigma_{x_j}$, then we set $x_j$ to true. Otherwise, $P'$ must pass through   $q^\sigma_{\overline{x_j}}$,
 and we set $x_j$ to false.
 In the following we replace each $q$-point by a small segment. The slabs of these segments will determine $\mathcal{A}^1$.  Consider an upward ray $r^1$ with positive slope starting at the $q$-point on $\lambda_{x_1}$, e.g., see Figure~\ref{graph}(c). Since all the edges that are currently in $\D_0$ have negative slopes, we can choose a sufficiently large positive slope for $r^1$ 
 and a point $a^1$ on $r^1$ such that all the slabs of $\D_0$ lie below $a^1$.
 We now find a point $b^1$ above $a^1$ on $r^1$ with sufficiently large $y$-coordinate
 such that the slab of $t_{c_1}b^1$ does not intersect the edges in $\D_0$.  
 Let $l^1_{\overline{x_1}}$ be the line determined by $r^1$. 
 For each $x_j$ and $\overline{x_j}$  (except for $j=1$), we now construct the lines  $l^1_{\overline{x_j}}$ and $l^1_{x_j}$  that pass through their corresponding $q$-points and intersect $r^1$ above $b^1$.
 The lines $l^1_{x_j}$ and $l^1_{\overline{x_j}}$ determine the arrangement $\mathcal{A}^1$. 
 Observe that one can construct these lines in the decreasing order of 
 the $x$-coordinates of their $q$-points, and ensure that 
 for each  $l^1_{x_j}$ ($l^1_{\overline{x_j}}$), there exists an interval $\lambda^{1}_{{x_j}}$
 ($\lambda^{1}_{\overline{x_j}}$) on the upper envelop of $\mathcal{A}^{1}$.
 Note that the correspondence is inverted, i.e.,  in $\mathcal{A}^1$, $\lambda^{1}_{\overline{x_j}}$ corresponds
 to $\lambda^0_{x_j}$, and  $\lambda^{1}_{x_j}$ corresponds to $\lambda^{0}_{\overline{x_j}}$.
 
For each   $j$, we draw a small  segment 
   $s^0_{x_j}$ ($s^0_{\overline{x_j}}$) perpendicular to $l^1_{x_j}$ ($l^1_{\overline{x_j}}$)
   that passes through the $q$-point and lies to the left of $q$, e.g., see  Figure~\ref{graph}(d).  
 The construction of $\D_i$, where $i>1$, is more involved. 
 The upper envelope of $\mathcal{A}^{i+1}$ 
 is determined by the upper envelope of the   slabs of the $s$-segments in  $\D_{i-1}$. For each $i$, we 
 construct the $q$-points and corresponding graph $G_{c_i}$. Appendix C includes the formal details. 

In the reduction we show that  any increasing-chord path $P$ from $t$ to $t'$ contains the points $t_{c_i}$.  We set a variable $x_j$ true or false depending on whether  $P$ passes through $s^0_{x_j}$ or $s^0_{\overline{x_j}}$.   The construction of $\D$ imposes the constraint that if $P$ passes through  $s^{i-1}_{x_j}$ ($s^{i-1}_{\overline{x_j}}$), then it must pass through $s^{i}_{x_j}$ ($s^{i}_{\overline{x_j}}$). Hence the truth values in all the clauses are set consistently. By construction of $G_{c_i}$, any increasing-chord path between $t_{c_{i-1}}$ to $t_{c_{i}}$ determines a satisfying truth assignment for $c_i$.  
 On the other hand, if $I$ admits a satisfying truth assignment, then for each clause $c_i$, we choose the corresponding increasing-chord path $P_i$ between $t_{c_{i-1}}$ and $t_{c_{i}}$. The union of all $P_i$ yields the required increasing-chord path $P$ from $t$ to $t'$. Appendix C presents the construction in details, and explains the challenges of encoding $\D$ in a polynomial number of bits.

\section{Open Problems}
\label{sec:con}
 The most intriguing problem in this context is to settle the computational complexity 
 of the increasing-chord path (\textsc{IC-Path}) problem. 
 Another interesting question is whether the problem \textsc{IC-Tree} remains NP-hard under the planarity
 constraint; a potential attempt to adapt our hardness reduction could be replacing the edge intersections by dummy vertices. 


\bibliographystyle{abbrv}
 
\bibliography{incchord}

\newpage

\section*{Appendix A}
\bigskip
\begin{minipage}{\linewidth}
\centering
\includegraphics[width=.8\textwidth]{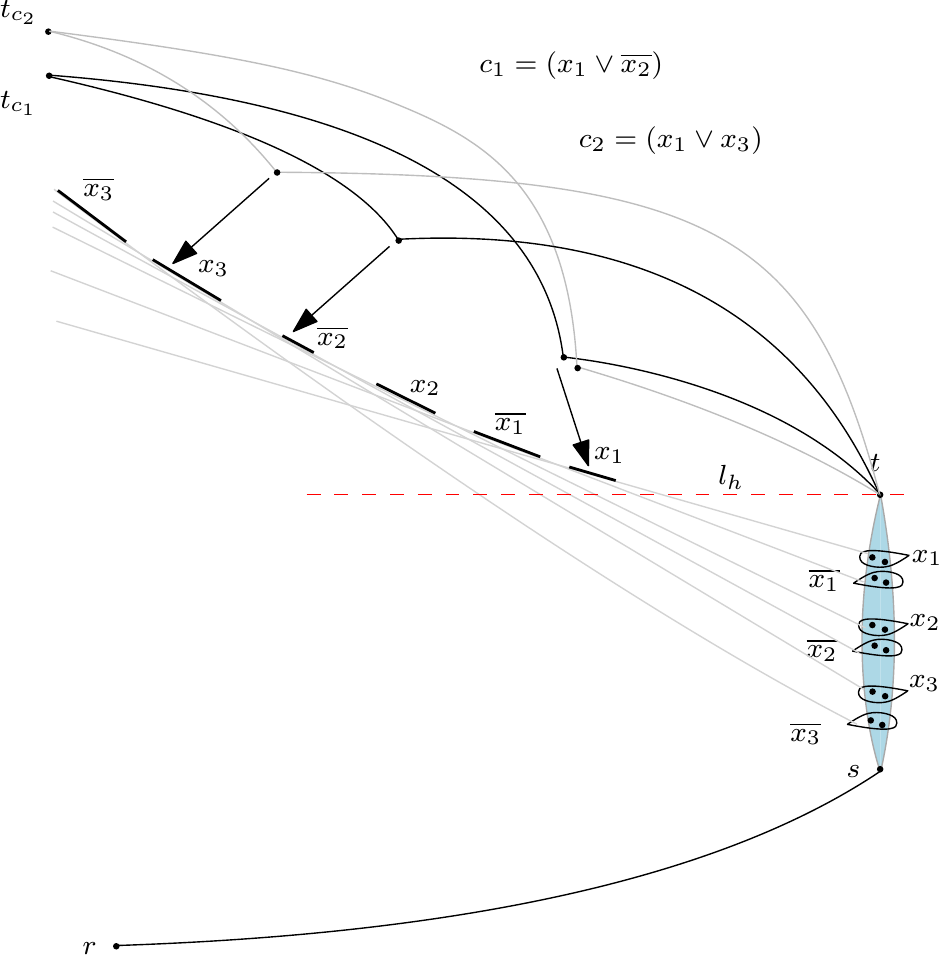} 
\captionof{figure}{Illustration for the hardness proof using a schematic representation of $\Gamma$. The points that correspond to $c_1$ and $c_2$ are connected in paths of  black, and lightgray, respectively.  The slabs of the edges of $H$ that determine the upper envelope are shown in lightgray straight lines. Each variable and its negation correspond to a pair of adjacent line segments on the upper envelope of  the slabs. 
}  
\label{idea}
\end{minipage}

\clearpage

\newpage
\noindent
\textbf{Lemma~\ref{lem:path}}
 \emph{Every increasing-chord path $P$ that starts at $s$ and ends at $t$ must pass through exactly one point among $p_{x_j}$ and $p_{\overline{x_j}}$, where $1 \le j\le \alpha$, and vice versa.}
\smallskip
\begin{proof}
By Observation~\ref{obs1},  $P$ must be $y$-monotone. Consequently, for each $j$, the edge on the $(2j)$th position on $P$ is a needle, which corresponds to either  $p_{{x_j}}p'_{{x_j}}$ or $p_{\overline{x_j}}p'_{\overline{x_j}}$. Therefore, it is straightforward to observe that $P$ passes through exactly one point among $p_{x_j}$ and $p_{\overline{x_j}}$. 

Now consider a path $P$ that  starts at $s$, ends at $t$,  and for each $j$,  
 passes through exactly one point among $p_{x_j}$ and $p_{\overline{x_j}}$. 
 By construction, $P$ must be $y$-monotone. 
 We now show that  $P$ is an increasing-chord path. Note that
 it suffices to show that for every straight-line segment $\ell$ on $P$,
 the slab of $\ell$ does not properly intersect $P$ except at $\ell$. 
 By Observation~\ref{obs1}, it will follow that $P$ is an increasing-chord path.

For every interior edge $e$ on $P$, which is not a needle, $e$ corresponds to some segment $\ell \in \{
p_{x_{j}}p'_{x_{j-1}},$ $ p_{x_{j}}p'_{\overline{x_{j-1}}}
p_{\overline{x_{j}}}p'_{x_{j-1}}, p_{\overline{x_{j}}}p'_{\overline{x_{j-1}}}
\}$, for some $1<j\le \alpha$. By construction, in each of these four cases,
 the needles incident to  $\ell$ lie either on the boundary or entirely outside of the
 the slab of $\ell$, and hence the slab does not properly intersect $P$ except at $\ell$.  
  Figures~\ref{precise}(a)--(b) illustrate the scenario when $\ell \in \{ 
p_{{x_{j}}}p'_{\overline{x_{j-1}}}, p_{{x_{j}}}p'_{x_{j-1}}
\}$.

Let $(s,a)$ and $(b,t)$ be the edges on $P$ incident to $s$ and $t$, respectively.
 By construction, these edges behave in the same way, i.e., all the needles on $P$ 
 are above the slab of $(s,a)$ and below the slab of $(b,t)$. 
 Consequently, the slab of $(s,a)$ (resp., $(b,t)$) does not properly intersect $P$ except at
 $(s,a)$ (resp., $(b,t)$).
 
For every interior edge $e$ on $P$, which is a needle, $e$ corresponds to some segment $\ell \in \{
p_{{x_j}}p'_{{x_j}},  p_{\overline{x_j}}p'_{\overline{x_j}}
\}$. By construction, the needles following (resp., preceding) $\ell$ on $P$ 
 are above (resp., below) the slab of $\ell$.  Consequently, the
  slab does not properly intersect $P$ except at $\ell$.  
  Figures~\ref{precise}(c)--(d) illustrate these scenarios.  
\end{proof}  

\newpage
\begin{minipage}{\linewidth}
\centering
\includegraphics[width=.5\textwidth]{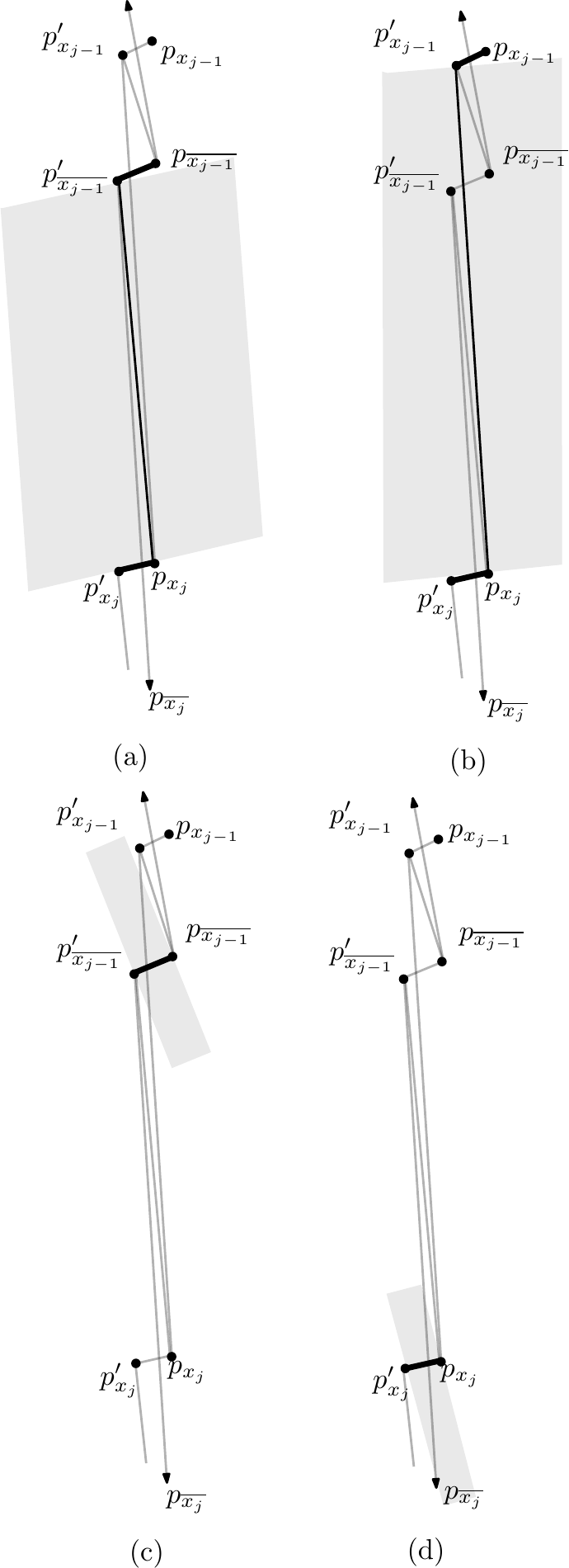} 
\captionof{figure}{Illustration for the slab of $\ell$. (a)--(b) The segment $\ell$ is not a needle.
(c)--(d) The segment $\ell$ is   a needle.
}  
\label{precise}
\end{minipage}


\newpage
\section*{Appendix B}
\begin{minipage}{\linewidth}
\centering
\includegraphics[width=.8\textwidth]{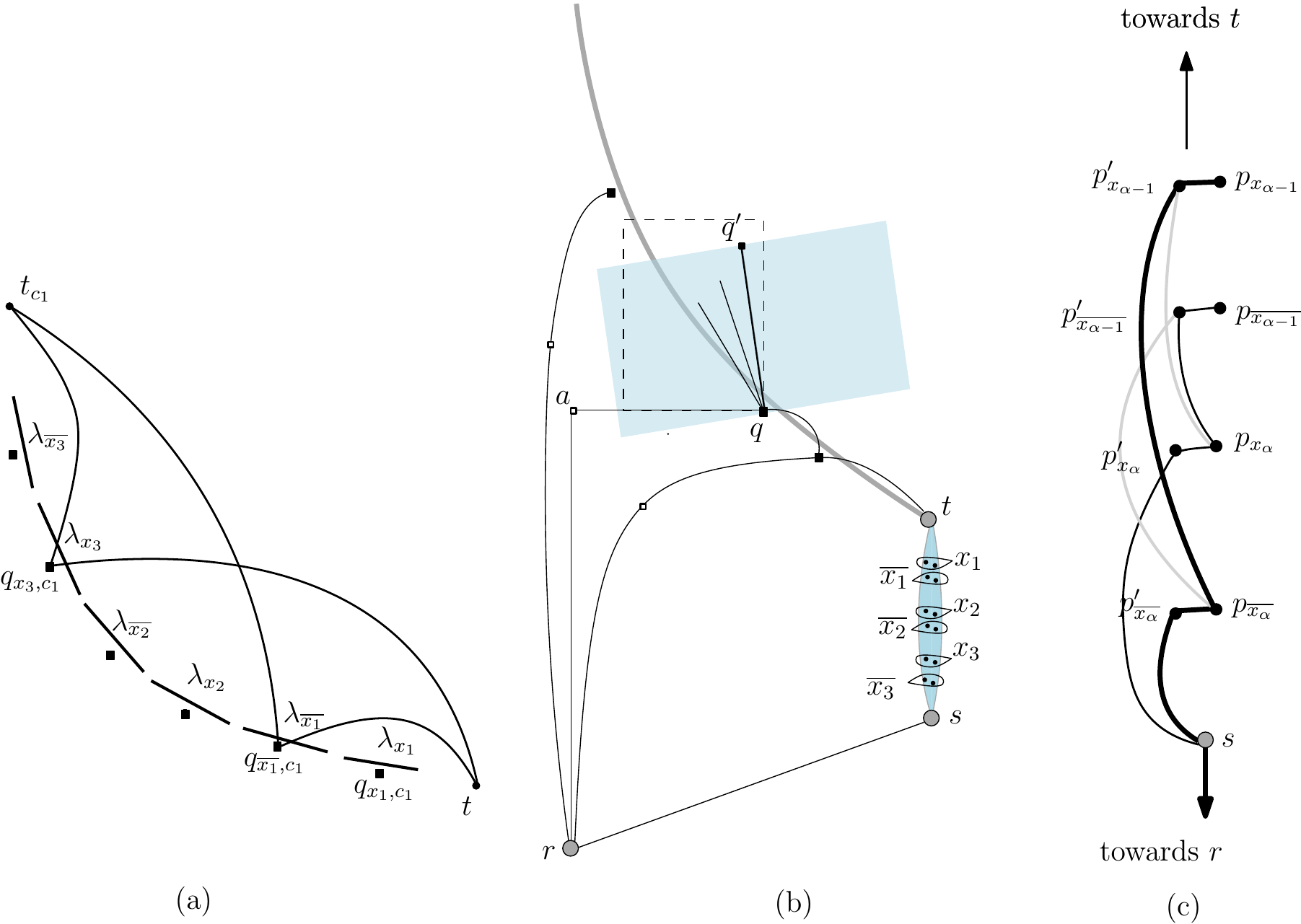} 
\captionof{figure}{(a) Construction of clause gadgets, where $c_1 = (\overline{x_1} \vee x_3)$.
 (b) A schematic representation of $\Gamma$. (c) Illustration for the reduction.
}  
\label{tree}
\end{minipage}
\clearpage
\newpage

Since determining whether a straight-line drawing of a tree  is an increasing-chord drawing is polynomial-time solvable~\cite{AlamdariCGLP12}, the problem \textsc{IC-Tree} is NP-complete. 
 We  now prove that $\Gamma$  admits an  increasing-chord rooted spanning tree if and only if $I$ admits a satisfying truth assignment.

{\textbf{Equivalence between the instances:}}    
 First assume that $I$ admits a satisfying truth assignment. We now construct an increasing-chord spanning tree $T$ rooted at $r$. We first choose a path $P$ from $r$ to $t$ such that it passes through either $p_{x_j}$ or $p_{\overline{x_j}}$, i.e., if $x_j$ is true (false), then we route the path through $p_{\overline{x_j}}$ ($p_{x_j}$). Figure~\ref{tree}(c) illustrates such a path $P$ in a thick black line, where $x_{\alpha}= \rm true$ and  $x_{\alpha-1}= \rm false$. 

Observe that only $2\alpha$ points remain below $l_h$, two points per literal, that do not belong to $P$. We connect these points in a $y$-monotone polygonal path $Q$ starting  at $s$, as illustrated in a thin black line in Figure~\ref{tree}(c). Note that $Q$ corresponds to a truth value assignment, which is opposite to the truth values determined by $P$. Therefore, by Lemma~\ref{lem:path}, $Q$ is also an increasing-chord path. Consequently, the point $t$ and the   points that lie below $l_h$ are now connected to $r$ through  increasing-chord paths. 

The tree $T$ now consists of the paths $P$ and $Q$, and thus does not span the vertices that lie above $l_h$. We now add  more paths to $T$ to span the points above $l_h$. Since every clause $c$ is satisfied,  
 we can choose a path  $P'$  from $t$ to $t_c$ that passes through 
 a literal-point  whose corresponding literal $x\in c$ is true.  Since  the  literal-points corresponding to
 true literals lie above  the slabs of $P$,   the path $P'$ determines an increasing-chord extension of $P$. Therefore, all the peaks and some literal-points above $l_h$ are now connected to $r$ via increasing-chord paths.  

For each remaining literal-point $q$, we add $q$ to $T$ via the increasing-chord path through its anchor. There are still some anchors that are not connected to $r$, i.e., the anchors   whose corresponding literal-points are already connected to $r$ via an increasing-chord extension of $P$. We connect each  anchor $a$ to $r$ via the straight line segment $ar$.

We now assume that $\Gamma$ contains an increasing-chord rooted spanning tree $T$, and show how to find  a satisfying   truth assignment for $I$. Since $T$ is rooted at $r$, and the peaks are not reachable via anchors, $T$ must contain an increasing-chord path $P = (r,s,\ldots,t)$ that for each variable $x_j$, passes through exactly one point among $p_{x_j}$ and $p_{\overline{x_j}}$. If $P$ passes through $p_{x_j}$  ($p_{\overline{x_j}}$), then we set $x_j$    to false (true).  Observe that passing through a variable $x_j$ or its negation selects a corresponding needle segment $p'_{x_j}p_{x_j}$ or $p'_{\overline{x_j}}p_{\overline{x_j}}$.  
 Recall that the interval $\lambda_{x_j}$ ($\lambda_{\overline{x_j}}$), which  corresponds to
  $p'_{x_j}p_{x_j}$ ($p'_{\overline{x_j}}p_{\overline{x_j}}$), lies above the literal-point
 $q_{x_j,c_i}$ ($q_{\overline{x_j},c_i}$), e.g.,
 see Figure~\ref{tree}(a).  
  Therefore, if the above truth assignment does not satisfy some clause $c$, then there cannot be any increasing-chord extension of $P$ that connects $t$ to $t_c$. Therefore, $T$ would not be a  spanning tree.

\section*{Appendix C}

Here we give the formal details of the construction of $\D$.

\textbf{Construction of $\D_0$:} 
 Choose $t_{c_0}$ ($t_{c_1}$) to be the bottommost (topmost)   point of $\lambda^0_{x_1}$ ($\lambda^0_{\overline{x_\alpha}}$). We then slightly shrink the intervals  
 $\lambda^0_{x_1}$ and $\lambda^0_{\overline{x_\alpha}}$ such that $t_{c_0}$ and  $t_{c_1}$ no longer belong to these segments.  If $c_1$  contains $\kappa$ literals, then there are $2^\kappa -1$ distinct truth assignment for its variables to satisfy $c_1$. For each satisfying truth assignment $\sigma_k$, where $1\le k \le 2^\kappa -1$, we construct a set of vertices and edges in $\D_0$, as follows.
 For each $x_j$ ($\overline{x_j}$), we construct a point $q^{\sigma_k}_{x_j}$ ($q^{\sigma_k}_{\overline{x_j}}$) at the midpoint of $\lambda^{i-1}_{x_j}$ ($\lambda^{i-1}_{\overline{x_j}}$). Later, we will refer to these points as \emph{$q$-points}, e.g., see Figure~\ref{graph}(c). 
 For each  $j$ from $1$ to $(\alpha-1)$, we make  $q^{\sigma_k}_{x_j}$ and $q^{\sigma_k}_{\overline{x_j}}$ adjacent to $q^{\sigma_k}_{x_{j+1}}, q^{\sigma_k}_{\overline{x_{j+1}}}$, e.g., see Figure~\ref{graph}(b). We then make $t_{c_0} $ ($t_{c_1} $) adjacent to the points corresponding to $x_1$ ($x_\alpha$) and its negation. Finally,  if  $x_j$ (resp., $\overline{x_j}$) is true in $\sigma_k$, then we remove the edges incident to $q^{\sigma_k}_{\overline{x_j}}$ (resp., $q^{\sigma_k}_{x_j}$).  We may assume that for each $x_j$, the points $q^{\sigma_k}_{x_j}$ lie at the same location. At the end of the construction, one may perturb them to remove vertex overlaps.

By Observation~\ref{obs1}, any $y$-monotone path $P'$ between $t_{c_0}$ and $t_{c_1}$ must be an increasing-chord path.  
 If $P'$ passes through  $q^\sigma_{x_j}$, then we set $x_j$ to true. Otherwise, $P'$ must pass through   $q^\sigma_{\overline{x_j}}$,
 and we set $x_j$ to false.
 In the following we replace each $q$-point by a small segment. The slabs of these segments will determine $\mathcal{A}^1$.  Consider an upward ray $r^1$ with positive slope starting at the $q$-point on $\lambda_{x_1}$, e.g., see Figure~\ref{graph}(c). Since all the edges that are currently in $\D_0$ have negative slopes, we can choose a sufficiently large positive slope for $r^1$ 
 and a point $a^1$ on $r^1$ such that all the slabs of $\D_0$ lie below $a^1$.
 We now find a point $b^1$ above $a^1$ on $r^1$ with sufficiently large $y$-coordinate
 such that the slab of $t_{c_1}b^1$ does not intersect the edges in $\D_0$.  
 Let $l^1_{\overline{x_1}}$ be the line determined by $r^1$. 
 For each $x_j$ and $\overline{x_j}$  (except for $j=1$), we now construct the lines  $l^1_{\overline{x_j}}$ and $l^1_{x_j}$  that pass through their corresponding $q$-points and intersect $r^1$ above $b^1$.
 The lines $l^1_{x_j}$ and $l^1_{\overline{x_j}}$ determine the arrangement $\mathcal{A}^1$. 
 Observe that one can construct these lines in the decreasing order of 
 the $x$-coordinates of their $q$-points, and ensure that 
 for each  $l^1_{x_j}$ ($l^1_{\overline{x_j}}$), there exists an interval $\lambda^{1}_{{x_j}}$
 ($\lambda^{1}_{\overline{x_j}}$) on the upper envelop of $\mathcal{A}^{1}$.
 Note that the correspondence is inverted, i.e.,  in $\mathcal{A}^1$, $\lambda^{1}_{\overline{x_j}}$ corresponds
 to $\lambda^0_{x_j}$, and  $\lambda^{1}_{x_j}$ corresponds to $\lambda^{0}_{\overline{x_j}}$.
 
For each   $j$, we draw a small  segment 
   $s^0_{x_j}$ ($s^0_{\overline{x_j}}$) perpendicular to $l^1_{x_j}$ ($l^1_{\overline{x_j}}$)
   that passes through the $q$-point and lies to the left of $q$, e.g., see  Figure~\ref{graph}(d).  
  We will refer to these segments as the \emph{$s$-segments}. 
     We choose the length of the $s$-segments small enough such that the slabs of these segments 
   still behave as lines of $\mathcal{A}^1$. 
   For each $s$-segment $qq'$, if there exists an edge $(w,q)$, where $w$ has a larger $y$-coordinate 
   than $q$, then then we delete the segment $wq$ and  add the line segment $wq'$.
   Since the slopes of the $s$-segments are negative,
   it is straightforward to verify that any $y$-monotone path between $t_{c_0}$ and $t_{c_1}$ will 
   be an increasing-chord path. 
   
This completes the construction of $\D_0$ and $\mathcal{A}^1$. 

\textbf{Construction of $\D_i$, where $i>0$:} 
  The construction for the subsequent drawing $\D_i$ depends on  $\mathcal{A}^i$,
  where $1\le i < \beta$, and the arrangement $\mathcal{A}^{i+1}$ is determined
  by $\D_i$. Although the construction of $\D_i$ from $\mathcal{A}^i$ is similar
  to the construction of $\D_0$ from $\mathcal{A}^0$, we need   
  $\D_i$ to satisfy some further conditions, as follows.

\begin{figure*}[pt] 
\centering
\includegraphics[width=.87\textwidth]{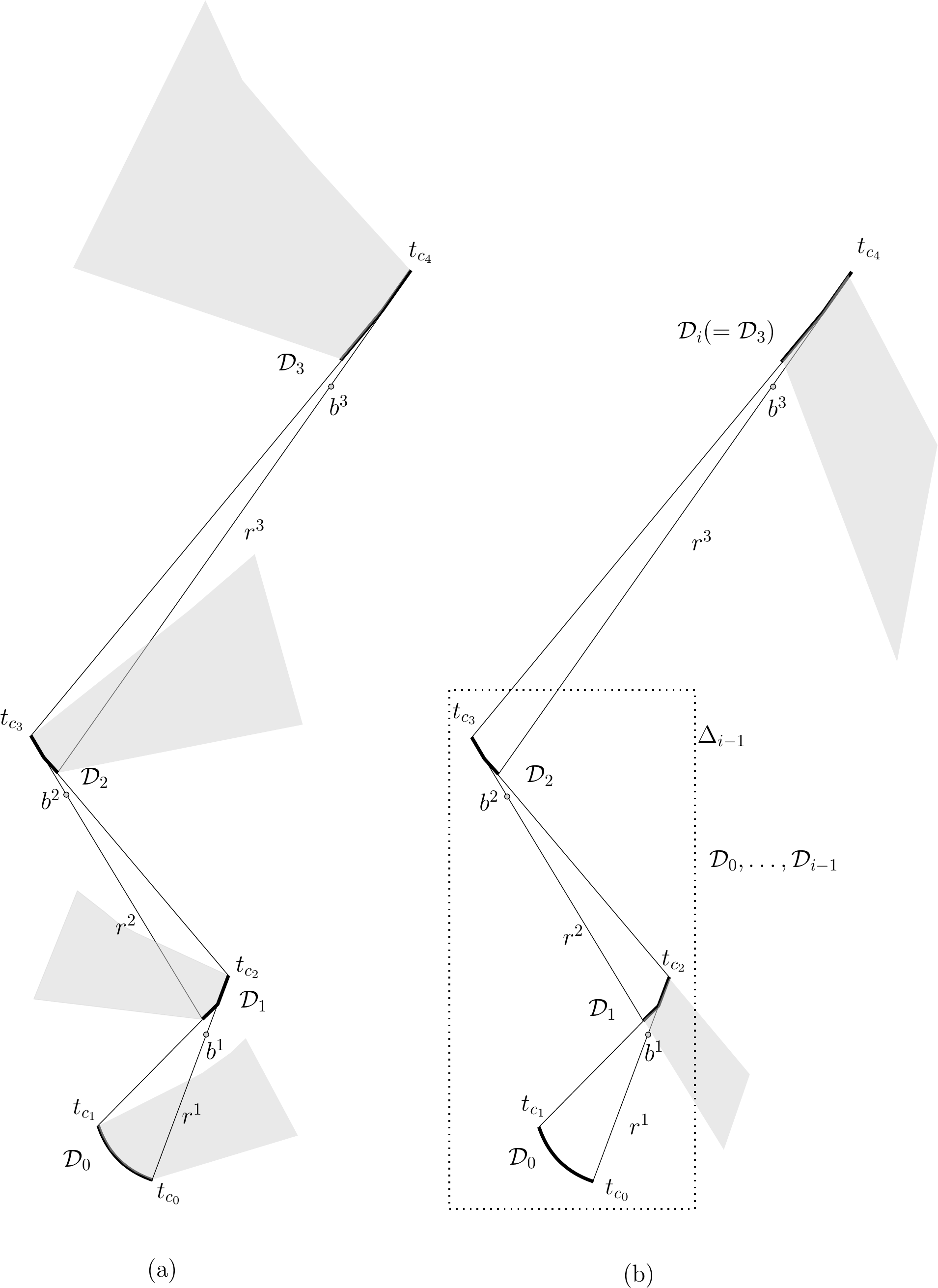} 
\caption{ (a) A schematic representation of $\D$. 
 The upward slabs of each $\D_i$ are illustrated in gray.
 (b) Construction of $\D_i$. The slabs of $\D_j$, where $j$ is odd,
   are shown in gray. Since the downward slabs of the edges 
  of $\D_i$ have positive slopes, and since the vertices of 
  $\D_i$ have larger $x$-coordinates than $\Delta_{i-1}$,
  the slabs of  $\D_i$ do not intersect $\D_0, \ldots,D_{i-1}$. 
}   
\label{overview}
\end{figure*}

 \begin{enumerate}
  \item[(A)] In $\mathcal{A}^{i+1}$, the segment $\lambda^{i+1}_{\overline{x_j}}$ (resp., $\lambda^{i+1}_{x_j}$) 
  plays the role of $\lambda^{i}_{x_j}$ (resp., $\lambda^{i}_{\overline{x_j}}$). 
  Therefore, the $q$-vertices and edges of $\D_i$ 
  must be constructed accordingly. 
  As a consequence, if an increasing-chord path $P'$ between $t_{c_{i-1}}$ and $t_{c_i}$ passes through 
  some $s^{i-1}_{x_j}$ ($s^{i-1}_{\overline{x_j}}$) in $\D_{i-1}$, then any increasing-chord extension of
  $P'$ to $t_{c_{i+1}}$ must pass through $s^i_{x_j}$ ($s^i_{\overline{x_j}}$) in $\D_{i}$.   
  
  \item[(B)] While constructing  $\D_i$, we must ensure that the slabs of the 
   segments in $\D_i$ do not intersect the segments in $\D_0,\ldots,\D_{i-1}$. 
   We now describe how to construct such a drawing $\D_i$, e.g., see Figure~\ref{overview}. 
   Without loss of generality assume that $i$ is odd. The construction when 
   $i$ is even is symmetric. Let $\Delta_{i-1}$ be the largest  
   $x$-coordinate among all the vertices in $\D_0,\ldots,\D_{i-1}$.
   
   Recall that the drawing of $\D_i$ depends on  $\mathcal{A}^i$, and we construct
    $\mathcal{A}^i$ starting with an upward ray $r^{i}$ and choosing a point 
    $b^{i}$ on $r^i$. We choose a positive slope for $r^i$, which is 
    larger than all the positive slopes determined by the slabs of $\D_0,\ldots,\D_{i-1}$.
    We then choose 
    $b^i$ with a sufficiently large $y$-coordinate such that  
    the $x$-coordinate of $b_i$ is larger than $\Delta_{i-1}$. It is now straightforward to 
    choose the lines of $\mathcal{A}^i$ such that their intersection points  
    are close to $b_i$, and have $x$-coordinates larger than $\Delta_{i-1}$. 
    Since the segments of $\D_i$ will have positive slopes, their slabs 
    cannot intersect the segments of $\D_0,\ldots,\D_{i-1}$.  
  \end{enumerate}

{\textbf{On the size of vertex coordinates:}} 
  Note that $\D$ has only  a polynomial 
 number of vertices, and our incremental construction for $\D$ is straightforward
 to carry out in polynomial number of steps. Therefore, the crucial challenge  is to 
 prove whether the vertices in $\D$ can be expressed in a polynomial number of bits or not. 
 
\begin{figure*}[h] 
\centering
\includegraphics[width=.87\textwidth]{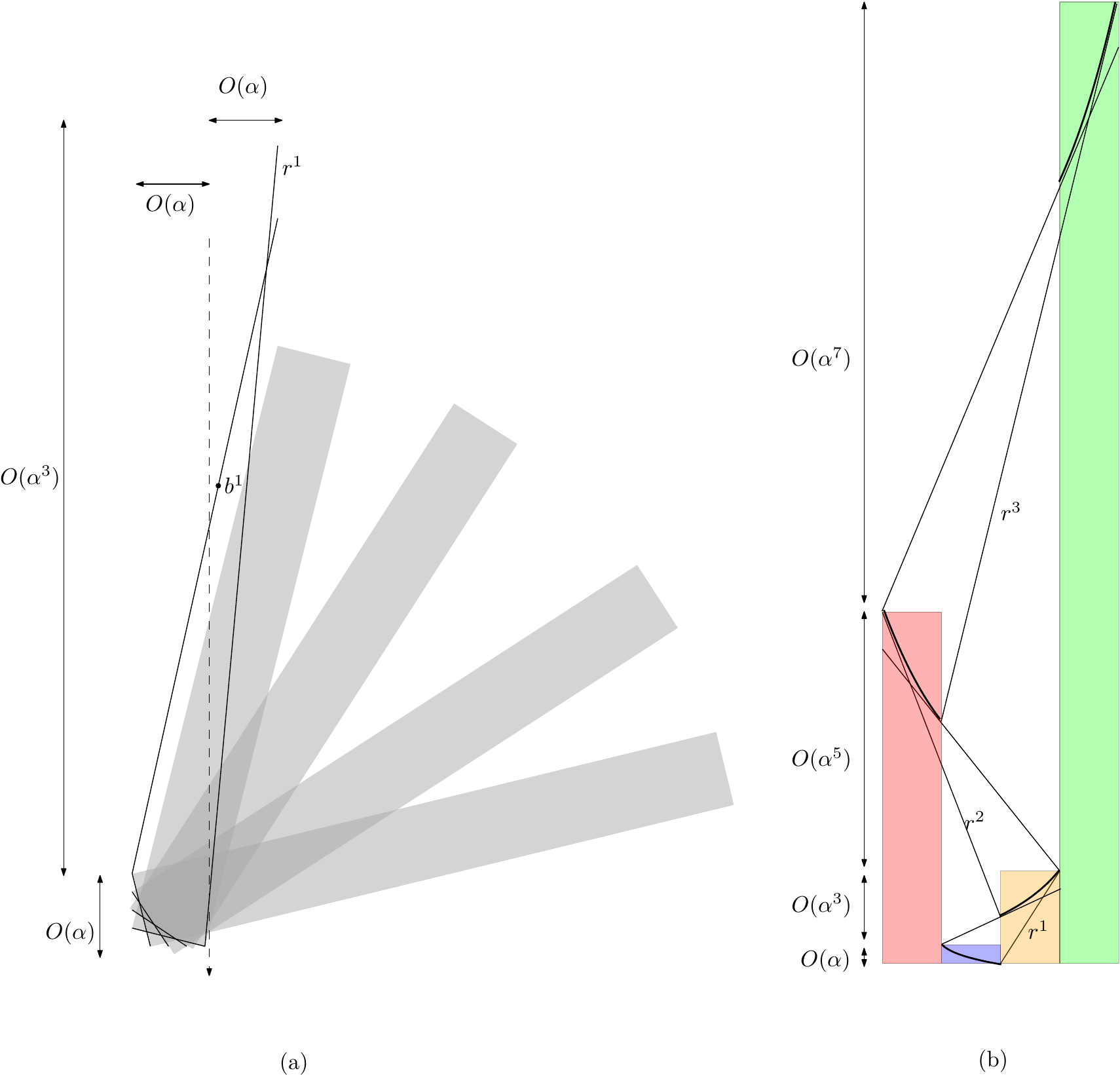} 
\caption{ (a) Construction of $r^1$. (b) Construction of $r^i$.
}   
\label{poly}
\end{figure*}

 As explained in the description of the construction of $\D_0$, 
 observe that the width and height of $\D_0$ is $O(\alpha)$, where $\alpha = |X|$.
 To construct $\D_1$, the crucial step is to choose the slope for $r^{1}$ and the point 
 $b^{1}$ on $r^1$. Since the largest slope of the slabs of $\mathcal{A}^1$ is $O(\alpha)$,
 it suffices to choose a slope of $\alpha^3$ for $r^1$. It is then straightforward  to choose a point 
 on $r^1$ as $b^1$, where $x$ and $y$-coordinates of $b^1$ are 
 of size $O(\alpha)$ and $O(\alpha^3)$, respectively, e.g., see Figure~\ref{poly}(a).
 Similarly, to construct $\D_i$ from $\D_{i-1}$, we can choose a slope of $\alpha^{2i+1}$ 
 for $r^{i}$, as illustrated in  Figure~\ref{poly}(b).
 Consequently, after $\beta$ steps, where $\beta = |C|$, the width of the drawing becomes
 $O(\alpha \cdot \beta)$ and the height becomes $\alpha^{O(\beta)}$. Note that we can describe  
 $r^i$ and $b^i$ in $O(\beta \log \alpha)$ bits.  However, encoding of the rest of
 the drawing seems difficult. For example, one can attempt to construct the 
 remaining vertices and edges of $\D$ using $r^i$ and $b^i$, as follows.

\begin{figure}[h] 
\centering
\includegraphics[width=.3\textwidth]{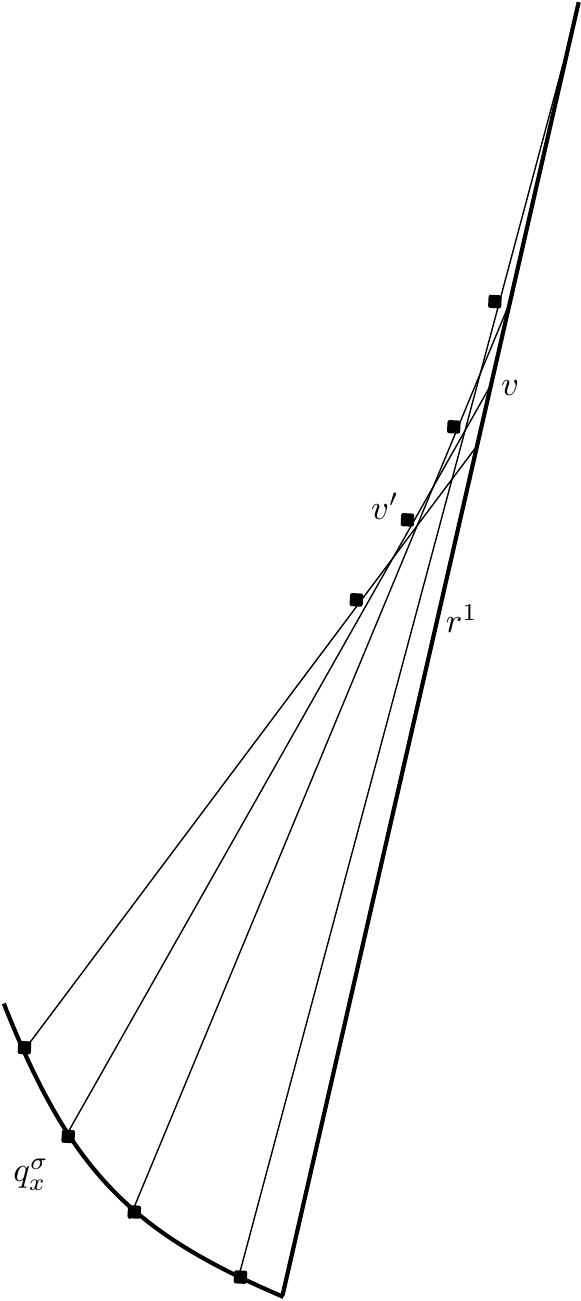} 
\caption{Illustration for the variables to encode the linear constraints.
}   
\label{lp}
\end{figure}

 Consider the upper envelope $U(\mathcal{A}^0)$ of $\mathcal{A}^0$. 
 To construct the straight lines for $\mathcal{A}^1$, one can construct 
 a set of variables and express the necessary constraints in a non-linear programming. Specifically, for each 
 $q$-point $q^\sigma_x$ on $U(\mathcal{A}^0)$, we create a variable point $v$
 on $r^1$  above $b^1$, and a variable $v'$ on the line
 $vq^\sigma_x$ on $U(\mathcal{A}^1)$, where the variable 
 $v'$ corresponds to the $q$-point on $U(\mathcal{A}^1)$.
 An example is illustrated in Figure~\ref{lp}.
 Similarly, for each $i>1$, we can create variables for the $q$-points of
 $\mathcal{A}^i$ from $\mathcal{A}^{i-1}$, $r^i$ and $b^i$.
 Since the number of vertices and edges of $\D$ is polynomial,
 the constraints we need to satisfy among all these $q$ points
 is polynomial. However, the solution size of such a nonlinear 
 system may not be bounded to a polynomial number of bits.

{\textbf{Equivalence between the instances:}} 
Any increasing-chord path $P$ from $t$ to $t'$ contains the points $t_{c_i}$.  We set a variable $x_j$ true or false depending on whether  $P$ passes through $s^0_{x_j}$ or $s^0_{\overline{x_j}}$.  By Condition (A), if $P$ passes through  $s^{i-1}_{x_j}$ ($s^{i-1}_{\overline{x_j}}$), then it must pass through $s^{i}_{x_j}$ ($s^{i}_{\overline{x_j}}$). Hence the truth values in all the clauses are set consistently. By construction of $\D$, any increasing-chord path between $t_{c_{i-1}}$ to $t_{c_{i}}$ determines a satisfying truth assignment for $c_i$. Hence the truth assignment satisfies all the clauses in $C$. 

On the other hand, if $I$ admits a satisfying truth assignment, then for each clause $c_i$, we choose the corresponding increasing-chord path $P_i$ between $t_{c_{i-1}}$ and $t_{c_{i}}$. Let $P$ be the union of all $P_i$.  By construction of $\D$, the slabs of $P_i$ do not intersect $P$ except at $P_i$. Hence, $P$ is the required increasing-chord path  from $t$ to $t'$. 


\end{document}